\documentclass[twocolumn,amsmath,amssymb,pre,a4paper]{revtex4}
\usepackage{graphicx}
\usepackage{dcolumn}
\usepackage{bm}

\begin{document}

\title{High order stimulated Brillouin scattering in single-mode fibers with strong feedback}

\author{Assaf Ben-Bassat, Ariel Gordon and Baruch Fischer}
\affiliation{Department of Electrical Engineering
Technion---Israel Institute of Technology, Haifa 32000, Israel}
\email{fischer@ee.technion.ac.il}

\begin{abstract}
We present an experimental and theoretical study of cascaded high
order Stimulated Brillouin Scatterings (SBS) in single mode
fibers. It is shown that because of the back-scattering nature of
the process, feedback in the input port is needed for obtaining a
significant cascaded effect in nonresonant systems. We also
discuss similarities to nonlinear photorefractive processes.

\end{abstract}

\maketitle

\section{Introduction}
Stimulated Brillouin Scattering (SBS) has a long research history
as a basic phenomenon and as a tool in many contexts and
materials. We mention here two important connections to SBS that
were intensively studied in the last three decades. The first one
is the link to phase conjugation \cite{zeldovich}. It was found
that the reflection of focused light beams in various media gave
in some cases a phase conjugate replica of the input beam. This
method gave maybe the first demonstration of phase conjugation,
and later generated many activities. Another wave of research was
related to SBS in fibers \cite{agrawalbook}, first in multi-mode
fibers, and then more intensively to single mode fibers. Here the
conventional meaning of phase conjugation in the spatial or
pictorial aspect is meaningless, especially when we consider
single mode fibers. Nevertheless, the nonlinear coupling
efficiency becomes very high even for low light powers at the mW
regime, due to the light confinement along large distances in the
fiber, compared to limited focused lengths that can be obtained in
free space propagation. Therefore, SBS became a crucial factor
that has to be considered in fiber-optic communications. It is
usually an effect that must be eliminated to allow the light
propagation without losing a big fraction of it to reflections. It
is similar to another effect that was used for phase conjugation,
the stimulated Raman scattering (SRS). In the SRS case however,
there were found important uses in fiber-optic communications. The
main one is the use as broadband amplifiers, especially at the
important $1.5 \mu m$ wavelength regime. A difference between SRS
and SBS is the magnitude of the frequency shift of the reflected
light (Stokes wave) compared to the input, originating from the
vibration frequency of the relevant medium entity involved in the
nonlinear process. This frequency shift is in fibers on the order
of $~10 GHz$ for SBS and $~13THz$ in fibers for SRS at the $1.5\mu
m$ wavelength regime.

In this paper we focus our attention on a cascaded SBS process in
single-mode fibers. Therefore the present work doesn't offer any
direct use for phase conjugation. Nevertheless, it can be
meaningful for other SBS schemes, in free space and multi-mode
fibers, where "spatial" phase conjugation is applicable.
Additionally, one might find possible applications by using the
self frequency shifts that are in the order of future dense WDM
(wavelenght division multiplexing) technologies, believed to be
heavily used in future fiber-optic communications.

For this paper, presented in the context of works on dynamic
holography and photorefractive optics, it is worthwhile to mention
some similarities between SBS, SRS and photorefractive four-wave
mixing. Pioneering work was done at the early stages when
photorefractive materials have been started to be a part of the
field of nonlinear-optics, and was used for wave mixing, phase
conjugation and oscillators, at a few places around the world: in
Kiev \cite{Kiev} (by Kukhtarev, Markov, Odulov, Soskin, and
Vinetskii), at Thomson CSF \cite{Huignard} (by Huignard, Spitz,
Aubourg, and Herriau, at the University of Southern California
\cite{Feinberg} (by Feinberg and Hellwarth), and at Caltech
\cite{PR0,PR1} (by Cronin-Golomb, Fischer, White and Yariv. Later,
a huge stream of research was done around the world in many
aspects of photorefractivity. We mention a few works on
photorefractive wave-mixing, done in our group at Technion
\cite{PR2,PR3,PR4,PR5,PR6,PR7,PR8,PR9} (by Fischer, Sternklar,
Weiss and Segev), that can be associated to the present work on
SBS. The first link is to a class of self oscillation processes in
photorefractive media \cite{Feinberg,PR1,PR2}. Like in SBS,
passive or self-pumped phase conjugate mirrors can be obtained.
Here four-wave mixing \cite{PR3,PR4} gives spontaneous phase
conjugate reflection via pump beams which are self generated and
can be regarded as the "internal" crystal "waves" (albeit light
waves), like the self generated "sound waves" in the SBS case
\cite{zeldovich}. The phase conjugation property can be also
explained by similar arguments, that among all possible
scattering, the phase conjugate pattern which is an oppositely
propagating replica of the input light wave (and therefore
coincide in space), experiences the highest gain, and thus wins
out and prevails over all other scatterings
\cite{zeldovich,PR5,PR6}. Another similarity is the self frequency
shift of the reflection with respect to the input light beam
\cite{PR7,PR8,PR9}. In the photorefractive case, the shift is
typically in the $1-10^{-3}Hz$ region, depending on the
photorefractive buildup time constant, which is much slower that
the relevant nonlinear effect in the SBS and SRS cases.
Additionally, for photorefractive wave mixing one can also think
of cascaded self reflections in an open or closed cavity. Specific
examples can be two-beam coupling, via reflection gratings where
the beams are almost counter-propagating, or resonators that give
high order oscillations.

It is also worthwhile to mention connections of fibers to phase
conjugation. In fact, one of the first suggestions for methods of
phase conjugation and its uses dealt with the restoration of
images transmitted through multi-mode fibers \cite{PC1}. It was
proposed there to use nonlinear three-wave mixing to phase
conjugate the distorted image transmitted through a fiber, and
retransmit it through an identical fiber section, such that the
second propagation exactly cancels the phase distortion of the
first section. The idea was later demonstrated \cite{PC2} in a
single section fiber with a round-trip propagation in the same
fiber, because of the difficulty to get two identical multi mode
fibers. Another idea in the early stages of research that gained a
lot of recent attention in the fiber-optic communication community
was to compensate for dispersion in single mode fibers by using
the phase conjugation property of flipping the spectral band of a
time dependent signal \cite{PC3}. Again a two section scheme with
phase conjugation between them, can provide a perfect
compensation.

SBS in fibers has been studied intensively throughout the years.
Input light at power levels on the order of 10 mW is strongly
backscattered, producing a frequency down shifted Stokes wave, due
to nonlinear interaction of light and sound waves. The associated
threshold depends on the light losses. The simplest configuration
for studying SBS in optical fibers is just a long enough optical
fiber, typically of a few km, with a good termination at its far
end, to avoid feedback. Much work has been done analyzing this
system, the SBS threshold \cite{feedback} and the reflection
strength, which is the ratio between the final power of the Stokes
wave and the initial power of the pump.

When feedback is added to the fiber from the far end termination,
or from other reflectors or simply by forming a ring cavity, the
SBS threshold can be lowered significantly and can even result in
oscillation and a Brillouin laser.

In a fiber with no feedback, SBS is well described by the common
"three wave model": the pump wave, the Stokes wave and the
mediating sound wave. Second order SBS \cite{2order}, which is the
generation of yet another, secondary Stokes wave, by SBS from the
first SBS wave, is known but considered to be weak for systems
without gain, and is usually neglected for such systems. However,
for systems with strong feedback, higher SBS orders can be
significant, and taking them into account is crucial for
understanding the physics of such systems.

In this work we investigate a system with strong feedback, where
several orders, a cascade of SBS, are generated, in a non-resonant
system (open cavity, with only one side feedback ). We realize
that it is necessary to put the feedback at the input port of the
fiber to allow the each SBS backscattering order to generate its
own SBS in an optimized intensity profile along the fiber
interaction path. We compare the experimental results to
theoretical analysis, and trace clearly the vast effects of second
and third order SBS. We find good agreement between the
experimental data and the multiple-order SBS theory.

\section{SBS without feedback}

In the simple Brillouin scattering scheme in long fibers most of
the energy of the input laser can be transferred to the Stokes
wave. Therefore one would expect that the Stokes wave would pass
its energy to a counter propagating Stokes wave moving again in
the direction of the input laser beam, and with a frequency of
$\omega_0-2\Omega$. Then one can ask if and what order can be
reached in such a chain of cascaded SBS system? (the n-th order
having an additional frequency shift, such that
$\omega_0-n\Omega$.) We will mathematically solve the coupled wave
equations for these three waves and then show experimentally that
for a regular system, even the generation of the second order
(third wave) SBS is negligible. Later we show that a chain process
that builds many strong high order SBS is possible by adding
feedback via a simple reflector at the input port.

The common three-wave SBS model (one acoustic and two light waves)
in steady state, is described by two coupled differential
equations, for the intensities of the pump and the Stokes wave
\cite{boydbook}:
$$ \frac{dI_{1}}{dz}=-gI_{1}I_{2} $$
\begin{equation}\label{I2}
  \frac{dI_{2}}{dz}=-gI_{1}I_{2},
\end{equation}
where $I_{1}$ and $I_{2}$ are the intensities of the incident and
the Stokes waves respectively. $g$ is the Brillouin gain
parameter, that depends on the fiber. These equations neglect
losses in the fibers and can be integrated analytically (as well
as the ones with nonzero losses \cite{Chen}) to yield the
following intensity profiles:
\begin{eqnarray}\label{IntI2}
  I_{2}(z)=\frac{I_{2}(0)(I_{1}(0)-I_{2}(0))}{I_{1}(0)e^{gz(I_{1}(0)-I_{2}(0))}-I_{2}(0)}
  \cr \cr  I_{1}(z)= I_{1}(0)-I_{2}(0)+I_{2}(z)
\end{eqnarray}
We choose our origin $z=0$ at the pump input port of the fiber.
Then $I_{1}(0)$ is the intensity of the incident wave, and
$I_{2}(0)$ is the output intensity of the Stokes wave.

It is seen that if  $I_2(L)$, the intensity of the Stokes wave at
the far side of the fiber, is zero, then $I_2(z) \equiv 0$, which
means that there is no Stokes wave at all. This reflects the fact
that the Stokes wave must start from a seed, whose source is noise
in the system. According to several approaches \cite{boyd1,boyd2},
it is important to think of the noise as distributed all over the
fiber. We shall follow here the simple case of a noise seed at the
far side of the fiber, on the order of 1nW.

Requiring $I_{2}(L)=\varepsilon$ in (\ref{IntI2}) yields a
relation between the incident power $I_1(0)$ and the power of the
Stokes wave $I_2(0)$ in terms of a transcendental equation. We
shall denote this relation by
\begin{equation}\label{B}
I_2(0)=B(I_1(0)).
\end{equation}
The function $B$ can be approximated, for low input intensities,
by
\begin{equation}\label{BB}
I_2(0)=\varepsilon e^{g L I_1(0)}
\end{equation}
This approximation is good as long as $\varepsilon e^{g L
I_1(0)}<<I_1(0)$ i.~e. for small enough incident intensities
$I_1(0)$, and can be obtained as well by the non-depleted pump
approximation.

Equations (\ref{I2}) neglect losses in the fiber. One of the
outcomes of losses in the fiber is the existence of a threshold
for SBS. It starts only if the incident beam is intense enough.
Otherwise, losses suppress the Stokes wave. Moreover, the transfer
of energy from the incident beam to the Stokes wave lasts only
while the intensity of the incident wave remains above the
threshold. Losses are also known to shorten the effective length
of the fiber, so the physical length in \ref{BB} is replaced by an
effective length \cite{agrawalbook}
\begin{equation}\label{leff}
L_{eff}= \frac 1 \alpha (1-e^{-\alpha L}),
\end{equation}
where $\alpha$ is the fiber loss coefficient.

Second order SBS requires a three optical wave model. The coupled
wave equations for the intensities are given by
\cite{2order,boydbook}:

\begin{eqnarray}\label{3waves}
  \frac{dI_1}{dz}&=&-gI_1I_2\cr\cr\frac{dI_2}{dz}&=&-gI_1I_2+gI_2I_3\cr\cr\frac{dI_3}{dz}&=&gI_2I_3
\end{eqnarray}

$I_{1}$ the incident wave, $I_{2}$ the back scattered Stokes wave,
$I_{3}$ the Stokes wave generated by $I_{2}$, which propagates in
the same direction as $I_1$, etc. For three optical waves the
equations can be integrated analytically \cite{2order}, but
unfortunately the three integration constants appearing in the
solution are again transcendental functions of the boundary
conditions. One can easily verify that
\begin{equation}\label{C}
C_1=I_{1}- I_{2}+ I_{3}  \quad{\rm and}\quad C_2=I_{1} I_{3}
\end{equation}
are constants of motion. Then we can find that
\begin{equation}\label{I1}
I_{1}(z)=\frac{C_1}{2}+(\frac q 2) \frac{1+C_3 e^{-qz}}{1-C_3
e^{-qz}}; \quad q=\sqrt{C_1^2-4C_2}
\end{equation}
$C_3$ is the third integral, and $I_{2}(z), I_{3}(z)$ can be
obtained from \ref{I1} through \ref{C}

In spite of the similarity of the terms $gI_{1} I_{2}$ and $
gI_{2} I_{3}$ in \ref{3waves}, power exchange between $I_{1}$ and
$I_{2}$ is much more efficient than between $I_{2}$ and $I_{3}$.
The difference stems form the boundary conditions. For a system
without feedback $I_{3}$ has the initial value of $\varepsilon$ at
$z=0$ and grows as $z$ increases, whereas $I_{2}$ decays as $z$
increases, keeping their product small all the way to $z=L$.

For a system without feedback or gain, the second order SBS is
weak compared to the pump and to the first order SBS. Indeed, from
the second relation of Eq. \ref{C} one obtains
\begin{equation}\label{3waves1}
I_{3}(z)=I_{1}(0) \frac{\varepsilon}{I_{1}(z)}
\end{equation}
Since $I_{1}(0)$ and $I_{1}(z)$ are roughly of the same order of
magnitude, one concludes that the second order SBS $I_{3}(z)$ is
not high above the noise level $\varepsilon$.

\begin{figure}[ht]
  \centering
  \includegraphics[width=7.7cm]{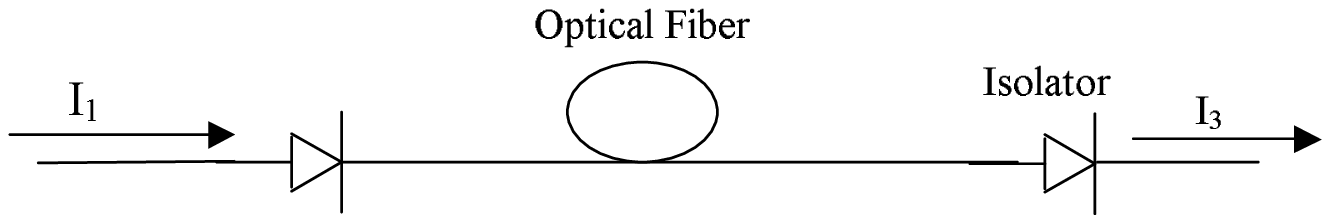}
  \caption{Experimental setup (regular - without feedback elements) for examining SBS .}\label{3setup}
\end{figure}

\begin{figure}[htb]
  \centering
  \includegraphics[width=7.7cm]{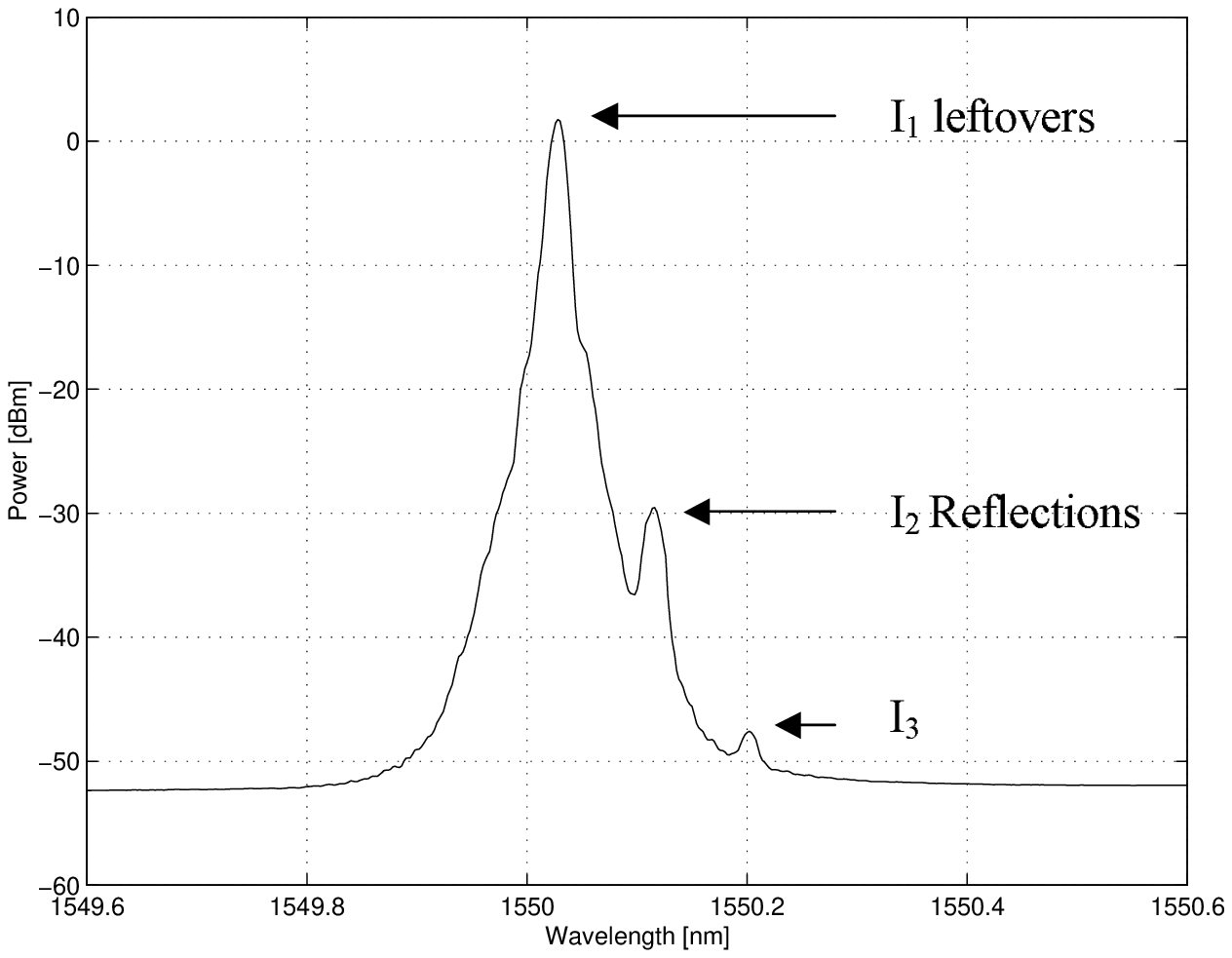}
  \caption{Optical spectrum of a simple SBS process.}\label{osa3}
\end{figure}

\begin{figure}[ht]
  \centering
  \includegraphics[width=7.7cm]{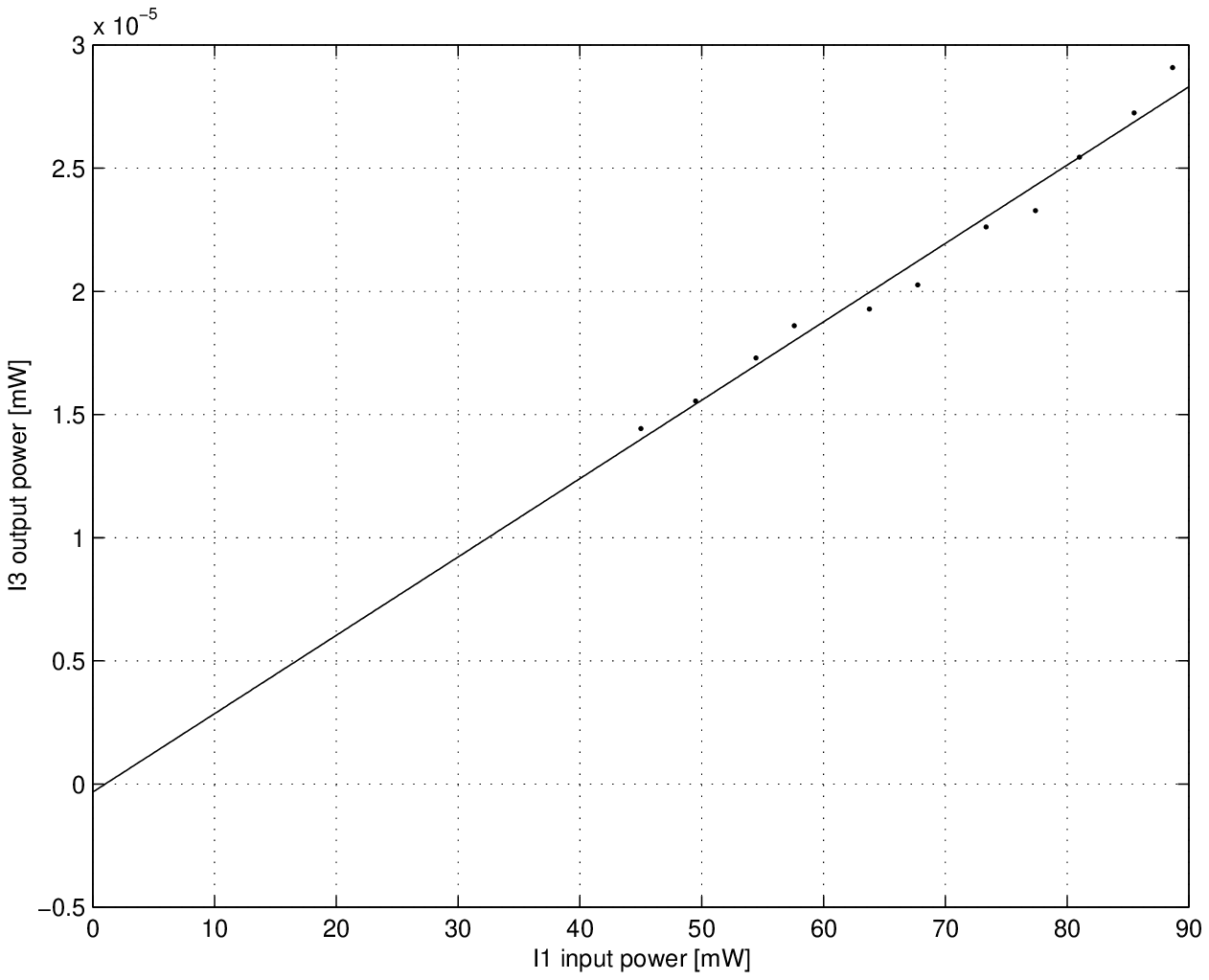}
  \caption{Output power of second stokes $(I_3)$ vs. the input power $(I_1)$ for a double SBS process without feedback.}\label{i3g}
\end{figure}

To verify Eq.(\ref{3waves1}), we measured $I_3$ using a very
simple experimental setup shown in Fig. \ref{3setup}. The fiber we
used was SMF-28, of 25 kilometers long. The output spectrum at
$Z=L$, shown in Fig. \ref{osa3}, is composed of three wavelengths,
$I_3(L),I_1(L)$ and the reflections from the input isolator of the
first Stokes wave $I_2$. In Fig. \ref{i3g} we have plotted
$I_{3}(L)$ vs. $I_{1}(0)$ . In the range of input intensities we
have applied, $I_{1}(L)$ was weakly dependent on $I_{1}(0)$ and
was about 1.5mW. We observe that $I_{3}(L)$ changes linearly with
$I_{1}(0)$ as expected from Eq. (\ref{3waves1}). The slope can be
related to $\varepsilon$ to yield $\varepsilon\approx 1nW$.

We thus summarize that a cascaded SBS beyond the first order SBS
without feedback elements is very weak. Nevertheless we show below
that with proper boundary conditions with one reflector, strong
higher order SBS can be generated.

\section{System with feedback and high order SBS }

\begin{figure}[h]
  \centering
  \includegraphics[width=7.7cm]{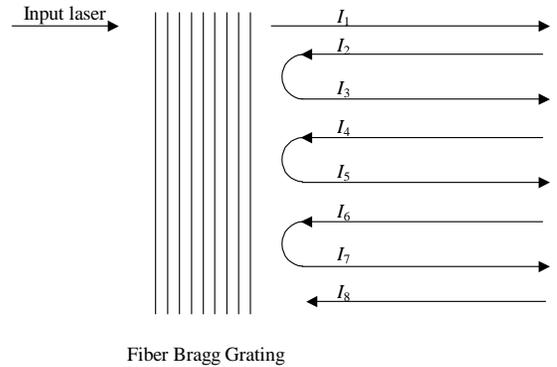}
  \caption{Cascaded SBS setup.}\label{multi}
\end{figure}

We have seen that the Stokes waves generated by SBS don't generate
their own Stokes waves because of the opposite growth direction
along the fiber of the "pump" and its SBS product. One way to
cascade many Brillouin scattered waves is by intervening in the
setup, causing every set of waves to resemble a two wave system.
Fig. \ref{multi} represents a suggested setup we check
experimentally. The input laser beam enters the system through a
fiber Bragg grating. The initial wave that start the cascading
process is obtained at the output of the grating. We will denote
this wave as $I_1$. It propagates to the right, generates a Stokes
wave $I_2$ that propagates to the left. We know that $I_2$ doesn't
generates its own Stokes, however when $I_2$ is reflected back
from the grating it will create a Stokes wave travelling again to
the left, since after the reflection the waves $I_3$ and $I_4$
behave according to the two wave system equations. $I_3$ begins
with a large power at the Bragg grating and is depleted only by
its Stokes wave $I_4$, which means that the coupled equations for
two waves can be used. This behavior is repeated for $I_3$ and
$I_4$, $I_5$ and $I_6$ and so on. From understanding how this
system works we can easily conclude that the coupling between
every pair of waves is only through the boundary conditions of
each pair. For the first pair the known boundary conditions are
given by $I_1(0)$ and $I_2(l_{eff})$, and for the second pair by
$I_2(0)$, which is the solution of the first pair, and by
$I_3(l_{eff})$ and so on. Each pair of waves gets it boundary
condition from the solution of the previous pair.

\begin{figure}[h]
  \centering
  \includegraphics[width=7.7cm]{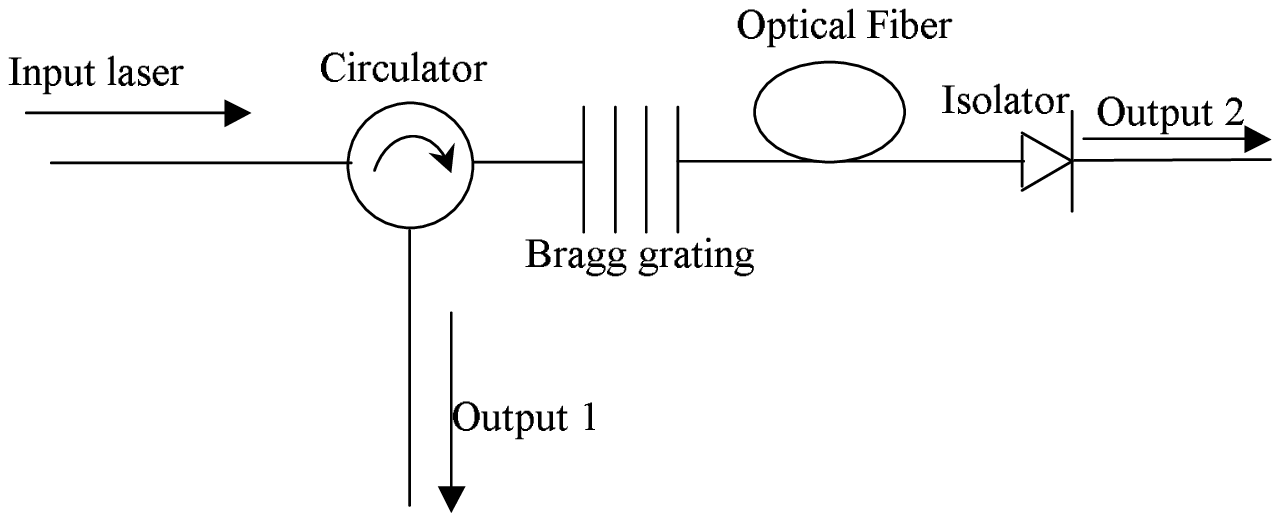}
  \caption{Experimental arrangement for multi stage SBS.}\label{multisetup}
\end{figure}

\begin{figure}[ht]
  \centering
  \includegraphics[width=7.7cm]{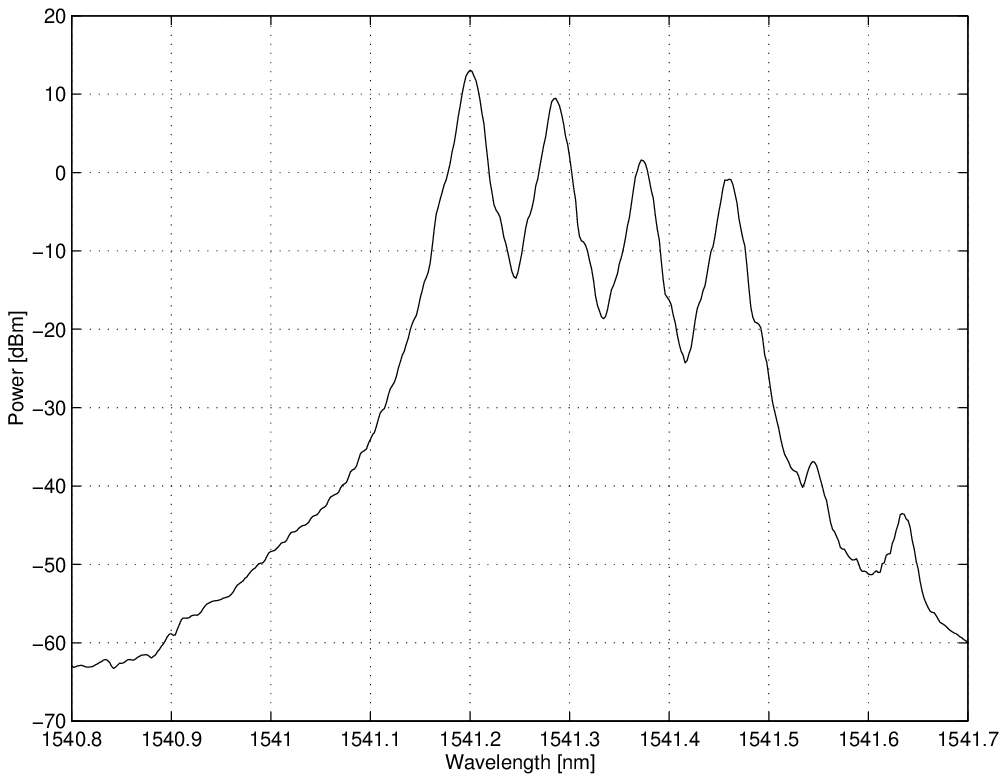}
  \caption{Output spectra of multi stage SBS measured at output 1.}\label{out1}
\end{figure}

\begin{figure}[h]
  \centering
  \includegraphics[width=7.9cm]{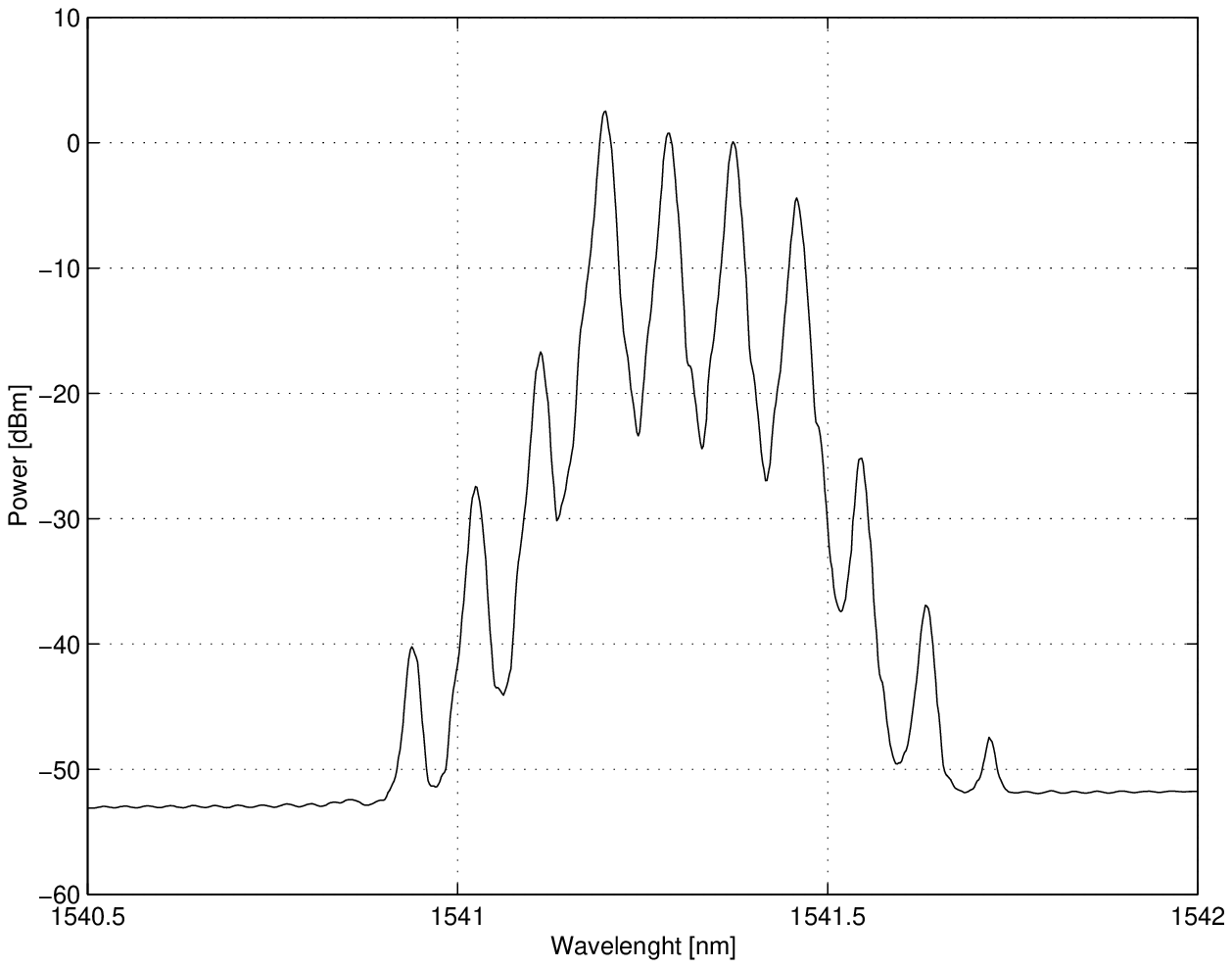}
  \caption{The experimental output spectrum of multi order SBS measured at the output port 2.}\label{out2}
\end{figure}

We first present the experimental result showing the generation of
strong high orders. The experimental setup is given in Fig.
\ref{multisetup}. We used a 25km long single-mode fiber. Since
every Stokes wave is down shifted by 10.3GHz from its "pump", we
used a broadband Bragg grating which can reflect all the Stokes
waves with approximately the same reflectivity. Knowing the input
power to our system and the reflection function of the grating is
sufficient to calculate the output spectra of both output1 and
output2. We expect to obtain at output port 2 a multiple-peaks
spectrum, with a 10.3GHz spacing, each of power $I_{th}$, since we
took a long enough fiber for the incident wave to be exhausted
down to the SBS threshold power. At output port 1 we should also
see a multiple-peaks spectrum. The first peak is due to the direct
reflection from the Bragg grating of the input laser with power of
$I_{in}\cdot r$, and all the rest are the back scattered Stokes
waves. In Figs. \ref{out1},\ref{out2} we show the output spectra
from ports 1 and 2. The central four strong lines belong to the
input and the first three SBS orders. We can see additional two
weaker lines at the right side giving the 4th and 5th SBS orders,
and at outpot port 2 additional three four-wave mixing products,
resulting mainly from the mixing of the input wave with the strong
first SBS orders.

For the theoretical part we write the coupled wave equations. We
solved numerically the equations for the first eight waves in the
system and compared them to the experiment.

The coupled equations for the eight waves are:
\begin{eqnarray}\label{8waves}
\frac{dI_1}{dz}&=&-gI_1I_2\cr\cr
\frac{dI_2}{dz}&=&-gI_1I_2+gI_2I_5\cr\cr
\frac{dI_3}{dz}&=&-gI_3I_4\cr\cr
\frac{dI_4}{dz}&=&-gI_3I_4+gI_4I_7\cr\cr
\frac{dI_5}{dz}&=&-gI_5I_6+gI_2I_5\cr\cr
\frac{dI_6}{dz}&=&-gI_5I_6\cr\cr
\frac{dI_7}{dz}&=&-gI_7I_8+gI_4I_7\cr\cr
\frac{dI_8}{dz}&=&-gI_7I_8
\end{eqnarray}

We note that the all SBS orders (even $n$ waves) are generated as
they propagate to the left direction, but then they are reflected
by the mirror at the left (input) side. This reflection enables
the cascaded process by generating the SBS. Thus the definition of
the eight wave used in the equations is as follows:\newline $I_1$:
frequency $\omega$ propagating to the right.\newline $I_2$:
frequency $\omega-\Delta\omega$ propagating to the left. \newline
$I_3$: frequency $\omega-\Delta\omega$ propagating to the right.
\newline$I_4$:
frequency $\omega-2\Delta\omega$ propagating to the left. \newline
$I_5$: frequency $\omega-2\Delta\omega$ propagating to the right.
\newline$I_6$:
frequency $\omega-3\Delta\omega$ propagating to the left. \newline
$I_7$: frequency $\omega-3\Delta\omega$ propagating to the right.
\newline$I_8$:
frequency $\omega-4\Delta\omega$ propagating to the left.
\newline

For the boundary conditions we have at the left side ($z=0$) the
reflectivity ratio $r$ between waves $I_3 \,\&\, I_2$, $I_5 \,\&\,
I_4$ and $I_7 \,\&\, I_6$, and at the right side ($z=L$) the
thermal noise ($I_2,I_4,I_6\& I_8$) needed for the SBS. Thus:
\begin{eqnarray}\label{bc}
&& I_{1}(z=0)=I_p \cr  && I_{n}(z=L)=\varepsilon \phantom{I_{n}}
\quad \, for \, n=2,4... \cr && I_{n}(z=0) = r I_{n-1} \quad \,
for \, n=3,5...
\end{eqnarray}

We don't elaborate here on the simulation results, that will be
given elsewhere, but note that they show a plausible match,
although not complete, to the experiments. There remains questions
and ingredients that have to be considered. An important  point is
the way that the seeding noise is incorporated into the system. In
a realistic model it should be taken as a stochastic source
distributed along the fiber. Additionally, other elements, such as
four-wave mixing and losses, should be included in some cases in
the calculations.

\section{Brillouin Laser}
\begin{figure}[h]
  \centering
  \includegraphics[width=7.7cm]{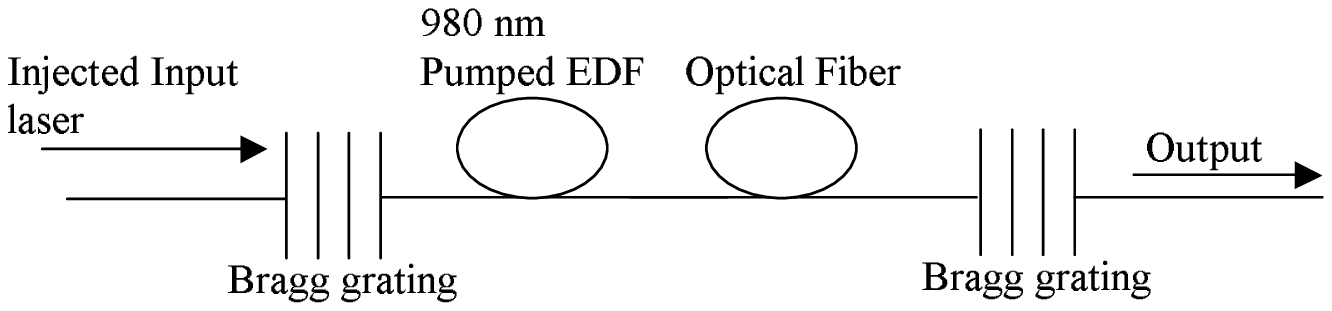}
  \caption{Experimental setup for a Brillouin laser.}\label{blaser}
\end{figure}

Understanding the simple cascading process for the multiple Stokes
waves can lead to a much more efficient setup for creating the
multi wavelength comb of Brillouin Stokes waves. The setup, shown
in Fig. \ref{blaser} is in the form of a long laser with feedback
in both sides of the cavity, and an external injected seed to
start the scattering process.
When the laser operates without the externally injected signal its
spectrum is governed only by the reflectivity spectrum of both
gratings, but when we start injecting an external laser source
through one of the gratings a process similar to the process in
the multi Stokes system happens and multi SBS Stokes appear. In
the laser configuration the Stokes waves have feedbacks on both
sides and the Stokes are travelling in an amplifying media,
therefore they are amplified inside the cavity which gives the
potential ability for many more Stokes waves. The  In this
configuration, of course, light is generated only in longitudinal
modes which meet the cavities longitudinal mode restriction, but
in a long cavity with relatively narrow spaced modes we see all
the Brillouin Stokes waves develop. In addition to the high order
SBS it is also possible to have products of four-wave mixing
(4WM). Every two SBS waves, can generate a new wave by 4WM. The
result are waves with the same frequency spacing, but here also
with a possible positive frequency shift; thus also getting new
lines with higher frequencies (or lower wavelengths).

\begin{figure}[t]
  \centering
  \includegraphics[width=7.7cm]{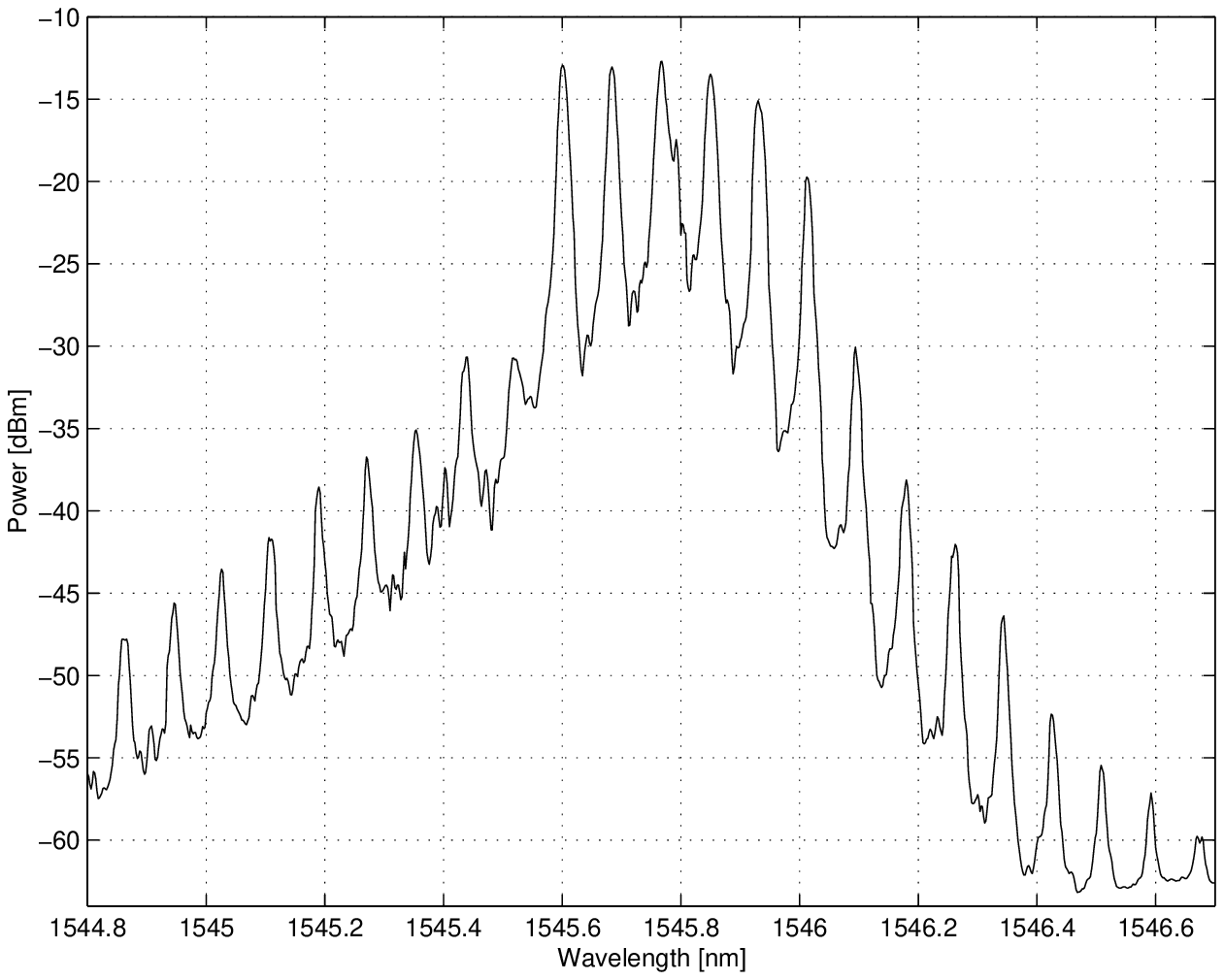}
  \caption{The experimental Brillouin laser output spectrum.}\label{Blaserout}
\end{figure}

In Fig. \ref{Blaserout} we show the output of the Brillouin laser
described above. Comparing the spectrum of the Brillouin laser to
that of the multi Stokes open system (Fig. (\ref{out2}) we see the
numerous number of Brillouin lines and also the lines generated by
four-wave mixing (4WM), especially those above the input
frequency.

In this experiment we used a chirped grating for one of the
'mirrors' and deliberately chose a grating that compensates for
the dispersion of one round trip in the laser, this selection
increased the amplitude of the 4WM terms compared to the same
laser with a non-chirped grating, but had no effect on the terms
created by SBS. We tested this view to show that SBS terms are
phased matched and that the peaks we see are mostly SBS and not
other non-linear phenomena.

\section{Discussion and Summary}

We have demonstrated that the SBS process does not cascade by
itself in open system configurations due to the power profile of
the waves in optical fibers. In order to cascade the SBS process
we must intervene in the system to change the basic configuration
of the interacting waves. One way of achieving this is by the use
of a Bragg reflector which changes the power profile to be
favorable for the generation of higher stokes reflection. In this
simple setup each pair of waves, signal and its Stokes, can be
treated as a simple SBS reflection and all pairs are related
through the boundary conditions of the setup. The boundary
condition relations make it simple to design the output power of
each Stokes wave by changing the input power and the Bragg
reflector's reflectivity. We have also demonstrated a closed
system of a laser configuration which is much more efficient than
the non-feedback setup and can generate many more SBS reflections,
but is not as simple to analyze and design.

\end{document}